\documentclass[journal]{IEEEtran}
\makeatletter
\def\ps@headings{%
\def\@oddhead{\mbox{}\scriptsize\rightmark \hfil \thepage}%
\def\@evenhead{\scriptsize\thepage \hfil \leftmark\mbox{}}%
\def\@oddfoot{}%
\def\@evenfoot{}}
\makeatother
\usepackage{setspace}
\usepackage{bm}
\usepackage{cite}
\usepackage{lettrine}
\usepackage{amssymb}
\usepackage{amsmath}
\usepackage{amsfonts}
\usepackage{enumerate}
\usepackage{indentfirst}
\usepackage{graphicx}
\usepackage{epstopdf}
\usepackage{multirow} %multirow for format of table
\usepackage{xcolor}
\usepackage{listings}
\usepackage{float}
\usepackage{subfigure}
\usepackage{amsthm}
\usepackage[top=0.75in, bottom=1in, left=0.65in, right=0.65in]{geometry}
\usepackage{algpseudocode}
\usepackage{algorithmicx,algorithm}
\usepackage{algorithm} 
\usepackage{url}

\newcommand{\beq}{\begin{equation}}
\newcommand{\eeq}{\end{equation}}

\newtheorem{proposition}{\bf Proposition}

\DeclareMathOperator*{\argmin}{argmin}

\makeatletter
\newcommand{\manuallabel}[2]{\def\@currentlabel{#2}\label{#1}}
\makeatother

\renewcommand{\Pr}{\mathrm{Pr}}

\newcommand{\tx}{\mathrm{Tx}}
\newcommand{\rx}{\mathrm{Rx}}

\newcommand{\PL}{P\! L}

\newcommand{\erf}{\mathrm{erf}}

\newcommand{\mm}{meta-material}
\newcommand{\Mm}{Meta-material}

\newcommand{\scenarioname}{meta-IoT sensing system}
\newcommand{\Scenarioname}{Meta-IoT sensing system}

\newcommand{\feavec}{received power vector}

\newcommand{\tsm}{temperature-sensitive material}

\newcommand{\hsm}{humidity-sensitive material}

\newcommand{\rcsf}{reflection coefficient function}

\newcommand{\mmr}{meta-IoT sensor}
\newcommand{\MmR}{Meta-IoT Sensor}

\newcommand{\mmu}{meta-IoT unit}

\newcommand{\mmrd}{meta-IoT structure}
\newcommand{\MmRD}{Meta-IoT Structure}

\newcommand{\sensfunc}{sensing function}
\newcommand{\SensFunc}{Sensing Function}

\newcommand{\rxunit}{processing unit}

\newcommand{\ist}{simultaneous sensing and transmission}

\newcommand{\dB}{\mathrm{dB}}
\newcommand{\err}{\mathrm{err}}

\begin{document}
\title{\huge{Meta-material Sensors based Internet of Things for \\6G Communications}}

\author{
\IEEEauthorblockN{
\small{Jingzhi Hu}\IEEEauthorrefmark{1},
\small{Hongliang~Zhang}\IEEEauthorrefmark{2},
\small{Boya~Di}\IEEEauthorrefmark{1},
\small{Kaigui~Bian}\IEEEauthorrefmark{3},
\small{and Lingyang~Song}\IEEEauthorrefmark{1}\\}
\IEEEauthorblockA{
	\IEEEauthorrefmark{1}\small{Department of Electronics, Peking University, Beijing, China,}\\
	\IEEEauthorrefmark{2}\small{Department of Electrical Engineering, Princeton University, Princeton, NJ, USA,}\\
	\IEEEauthorrefmark{3}\small{Department of Computer Science, Peking University, Beijing, China}\\}
\thanks{This work has been submitted to the IEEE for possible publication.
Copyright may be transferred without notice, after which this version may no longer be accessible.}
}

\maketitle

\setlength{\abovecaptionskip}{0pt}
\setlength{\belowcaptionskip}{-10pt}
\begin{abstract}
In the coming 6G communications, the internet of things (IoT) serves as a key enabler to collect environmental information and is expected to achieve ubiquitous deployment. 
However, it is challenging for traditional IoT sensors to meet this expectation because of their requirements of power supplies and frequent maintenance, which are due to their power-demanding sense and transmit modules. 
To address this challenge, we propose a meta-IoT sensing system, where the IoT sensors are based on specially designed meta-materials.
The meta-IoT sensors achieve simultaneous sensing and transmission by physical reflection and require no power supplies.
In order to design a meta-IoT sensing system with optimal sensing accuracy, we jointly consider the sensing and transmission of meta-IoT sensors and propose efficient algorithms to optimize the meta-IoT structure and the sensing function at the receiver. 
As an example, we apply the meta-IoT system to sensing environmental temperature and humidity levels. 
Simulation results show that by using the proposed algorithm, the sensing accuracy can be largely increased.
\end{abstract}

\section{Introduction}
% 在新的6G之中对于环境感知提出了新的需求, 也就是前两句需要换为对于IoT的需求.
For the next coming 6G communications, it is envisioned that the internet of things~(IoT) lays the foundation for various important sensing applications~\cite{Wikstrom2020Challenges}.
To support sensing applications in intelligent industrial processing and environmental monitoring, an extremely large number of IoT sensors need to be spread pervasively in the environments to collect information.
The number of IoT sensors needed in 6G is reckoned to be 10-fold more than that in 5G, reaching 10 million devices per square km~\cite{Liu2020Vision}.
% Liu2020Vision 是不能删除的
In order to support the ultra-massive deployment, it is necessary for the IoT sensors in 6G to have extremely low power consumption, so that they can be energy-saving and used for continuous sensing without any human intervention or maintenance for an ultra-long time~\cite{Wikstrom2020Challenges}.

% CITE Communications in the 6G era
% CITE Wikstrom2020Challenges
Nevertheless, it is challenging for existing IoT sensors to satisfy the demands of 6G.
Because existing IoT sensors need energy suppliers, such as lithium batteries or energy harvesters, to support their power-consuming sensing, modulation, and transmission modules.
Specifically, the power consumption and sophisticated microchips needed in the modulation of sensing results and transmission of the signals result in non-negligible costs and expenses, which make these IoT sensors not suitable for the ultra-massive deployment.
% 落脚点在sensing and communication上面.
To meet the demand for pervasive environment sensing in 6G, it is expected to develop sensors with \emph{simultaneous sensing and transmission}, where sensing and transmission are performed simultaneously through physical signal reflection.
By this means, no extra energy or sophisticated microchips are required.

Fortunately, meta-material sensors have shown the potential of {\ist} for sensing applications in 6G communications~\cite{Vena2014AFully, RN639}, which we refer to as the \emph{meta-IoT sensors}.
% RN639是不能删除的.
The {\mmr}s are printed circuits on supportive substrates combined with some sensitive materials, which together work as reflectors for wireless signals.
Their working principle is that their reflection coefficients for wireless signals are sensitive to surrounding environmental conditions.
Therefore, by analyzing the reflected signals from the {\mmr}s, the influence of the environmental conditions can be recognized, and the values of the environmental conditions can be estimated.

% 这里提出几个无线被动式的传感器
% 需要增加一句铺垫
In literature, several works have discussed using {\mm}s for sensing environmental conditions.
In~\cite{Vena2014AFully}, the authors proposed a {\mm} sensor which is composed of split-ring resonators~(SRRs) and a temperature-sensitive polymer, which can be used to sense CO$_2$ concentration or temperature.
% 这个文章中设计了一种传感器, 对于温度和二氧化碳浓度都敏感,
In~\cite{Lu2018ANovel}, the authors proposed a {\mm} temperature sensor with a double SRR, which can sense temperature levels in harsh high-temperature environments.
In~\cite{Ni2016Humidity}, the authors designed a humidity sensor based on a perfect {\mm} absorber.
% 这个文章使用了一个湿度的传感器
Moreover, in~\cite{Dong2018Applying}, the authors utilized {\mm}s to design an enhanced passive humidity sensor.

%However, above literature focuses on the {\mmr} design to improve the sensing performance,  while the influence of the integrated communication part on the climate sensing has not been considered jointly.
However, the above literature focuses on the design to improve the sensing performance while the transmission lacks joint consideration.
%Specifically, if the SNR of the signals reflected by the {\mmr} is low at the wireless receiver, the sensing accuracy will suffer.
As the design of {\mmr}s not only influences the sensing performance, but also has an impact on the signal transmission, it is important to design the {\mmr} joint considering both sensing and transmission.
%To handle this challenge, an efficient method to jointly optimize the {\mmr} structure and signal process function for the {\mmr} climate sensing application is in need.

% 在这里大概说一下我们这篇文章打算做什么. 我们是怎么样 handle 上面列举出的这些工作所无法考虑到的缺陷的.
% (而我们要做的是感知和传输的一体化)
%
%\pdfcomment[icon=Note, color=yellow]{Insert the followings in tex file!} 
% related work里面引用的不是通信和感知分离的传统传感器的文章, 而是以前的超材料传感器的文章. 这是由于: 传统的温湿度传感器由于和我们的文章差别太远, 作为相关文献并不合适, 此外也有若干已有的工作做温度或者湿度的超材料传感器, 因此这里选用了那些已有的、和本文更相关的文章作为related work. 
%这些related work, 也是被动式的无线反射传感器, 因此上面的那些文章也是现有的超材料传感器, 它们也算是sensing 和 communication的结合.
%因此在这里说的, 是我们的文章相比起已有的超材料传感器的不同, 即具有多功能以及提出了联合超材料设计和感知方程优化的问题与算法.

In this paper, we design a general meta-IoT sensing system, which is able to sense multiple environmental conditions.
% 这里要再次强调一下难点
%难点可以说是由于反射系数难以建模, 难以对超材料本身的结构进行优化, 并和{\sensfunc} 一起联合优化.
We jointly considering the influence of both sensing and transmission and formulate a joint {\mmrd} and {\sensfunc} optimization problem, which is solved efficiently through problem decomposition.
The simulation results verify the effectiveness of the proposed design to optimize the {\mmr} design in terms of sensing accuracy.

% 总结一下我们文章的贡献
The rest of the paper is organized as follows.
In Section~\ref{sec: mm sensor model}, we introduce the proposed {\mmr}s.
In Section~\ref{sec: system model}, the model of the {\scenarioname} is described.
In Section~\ref{sec: problem formulation}, we formulate a joint {\mmrd} and {\sensfunc} optimization problem and propose the algorithm to solve it in Section~\ref{sec: alg design}.
Simulation results are provided in Section~\ref{sec: simulation result}, and a conclusion is drawn in Section~\ref{sec: conclu}.

%%%%%%%%%%%%%%%
\section{{\MmR}s}
%%%%%%%%%%%%%%%
\label{sec: mm sensor model}

{\Mm}s are artificial periodic structures exhibiting exotic properties, which are underpinned by their special frequency responses for wireless signals~\cite{our_ris_work,ElMossallamy2020Reconfigurable}.
By designing a {\mmr} with specific structure and sensitive materials, we can make the frequency response of the {\mmr} sensitive to various sensing targets, such as temperature, humidity, gas concentration, and so on.

\begin{figure}[!t]
\center{\includegraphics[width=0.8\linewidth]{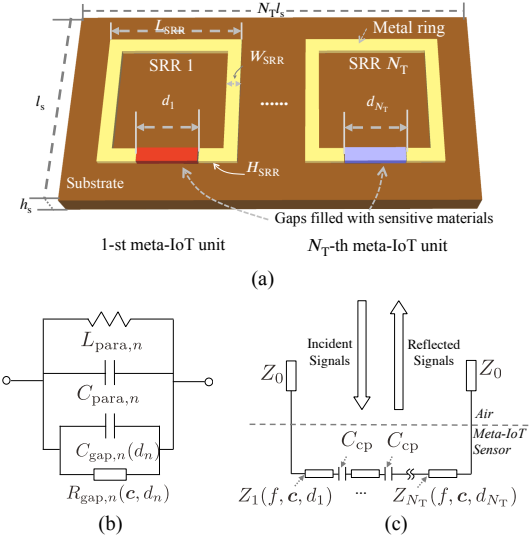}}
	\vspace{-.5em}
	\setlength{\belowcaptionskip}{-1.em} %调整图片标题与下文距离
	\caption{(a) A {\mmr} for sensing $N_\mathrm{T}$ different environmental conditions. (b) Equivalent circuit model of the $n$-th {\mmu}. (c) Equivalent circuit model of the {\mmr}.}
	\vspace{-1.em}
	\label{fig: sensor overview}
\end{figure}

As shown in Fig.~\ref{fig: sensor overview}~(a), the {\mmr} consists of $N_\mathrm{T}$ {\mmu}s for $N_\mathrm{T}$ different target environmental conditions, which we refer to as $N_\mathrm{T}$ \emph{sensing target conditions}.
Each {\mmu} consists of a SRR with a horizontal gap printed on a supportive sensitive substrate.
The substrate of the {\mmr} is made of dielectric materials, and the SRR is made of metal.
The detailed dimensions of a {\mmr} are illustrated in Fig.~\ref{fig: sensor overview}~(a).

Each {\mmu} can be approximated by a RLC resonant circuit as shown in Fig.~\ref{fig: sensor overview}~(b).
For the $n$-th {\mmu} with gap width $d_{n}$, given the \emph{sensing target condition vector} of the $N_\mathrm{T}$ sensing targets being $\bm c=(c_1, \dots, c_{N_\mathrm{T}})$, the impedance of the circuit can be calculated as\footnote{For the sake of generality, we consider that the electric properties of the $n$-th sensitive material can be influenced not only by the $n$-th sensing target, but also by the other sensing targets as well.}
% 思路: 符号替换
%R_{mat, n} -> R_{M, n}
%C_{surf, n} -> C_{S, n}
%C_{gap, n} -> C_{G, n}
\begin{align}
\label{equ: impendance of v-SRR}
Z_n(f, \bm c, d_{n})  \!=&\big( 
{1\over 2\pi\mathrm{i} f L_{\mathrm{para},n}} \!+\! 
{2\pi\mathrm{i} f C_{\mathrm{para},n}} \!+\! {2\pi\mathrm{i} f C_{\mathrm{gap},n}(d_n)}\nonumber \\
 &+ {1\over R_{\mathrm{gap},n}(\bm c, d_n)}
\big)^{-1},
\end{align}
where $\mathrm{i}$ denotes the imaginary unit, $f$ denotes the frequency of incident signals on the {\mmu}, 
and
$L_{\mathrm{para},n}$ and $C_{\mathrm{para},n}$ are the parasitic inductance and capacitance of the SRR, respectively.
Besides, $R_{\mathrm{gap},n}(\bm c, d_n)$ and $C_{\mathrm{gap},n}(d_n)$ can be modeled as
\begin{align}
\label{equ: resistance with d}
& R_{\mathrm{gap},n}(\bm c, d_n)  \!=\! \frac{d_n}{\rho_{\mathrm{mat}, n}(\bm c)W_{SRR}H_{\mathrm{SRR}}},
C_{\mathrm{gap},n}(d_n) \!=\! \frac{\hat{C}_{\mathrm{gap},n}}{d_n}, 
\end{align}
where $\rho_{\mathrm{mat}, n}(\bm c)$ denotes the conductivity of the $n$-th sensitive material when sensing target conditions being $\bm c$,
and $\hat{C}_{\mathrm{gap},n}$ denotes capacity of the gap with a unit width.
Then, as shown in Fig.~\ref{fig: sensor overview}~(c), the total impedance of the {\mmr} can be expressed as 
\beq
\label{equ: total impedance}
Z(f, \bm c, \bm d) \!=\! \big(\!
\sum_{n=1}^{N_\mathrm{T}}\! Z_n(f, \bm c,  d_n)^{-1} \!+\! {N_\mathrm{T}\!-\!1\over 2\pi\mathrm{i} f C_{\mathrm{cp}}}
\big)_,^{-1}
\eeq
where $C_{\mathrm{cp}}$ denotes the capacity due to the coupling between adjacent {\mmu}s, 
and $\bm d = (d_1, ..., d_{N_\mathrm{T}})$.

For the {\mmr}, its \emph{reflection coefficient} is a parameter that describes the fraction of the wireless signals reflected by an impedance discontinuity in the transmission medium~\cite{pozar2011microwave}.
%It can be calculated by the ratio between the electric field intensities of the incident and reflected signals.
In this paper, to facilitate the design of meta-IoT systems, we focus on the reflection coefficient which describe the ratio between the reflected and incident power.
% CITE Wireless communications with programmable metasurface: Transceiver design and experimental results
Based on~\cite{pozar2011microwave}, the reflection coefficient can be analytically modeled by 
\begin{equation}
\label{equ: reflection coefficient}
\hat{\gamma}(f,  \bm c,\bm d) = \Big|{Z(f,  \bm c, \bm d) - Z_0\over Z(f,  \bm c, \bm d) + Z_0}\Big|^2,
\end{equation}
where $Z_0=377~\Omega$ is the impedance of free space.

By substituting~(\ref{equ: impendance of v-SRR}), (\ref{equ: resistance with d}), and (\ref{equ: total impedance}) into~(\ref{equ: reflection coefficient}), we can observe that the reflection coefficient of the {\mmr} is dependent on $\bm d$.
Therefore, the gap widths of the $N_\mathrm{T}$ {\mmu}s can be considered as the variables to design the {\mmr}, which are thus referred to as the \emph{{\mmrd} vector}.

% 基于这个提出的协议, 我们设计了优化问题, 在其中感知和通信都得到了考虑, 并且基于他们去设计.
Using the analytical model derived above, i.e., $\hat{\gamma}(f,  \bm c,\bm d)$, we reveal the influence of $\bm d$ on the reflection coefficients.
%Specifically, the {\mmrd} vector determine the impedance of the SRRs in each {\mmu} and the reflection coefficient of the complete {\mmr}.
%Besides, in~(\ref{equ: resistance with d}), it can be observed that by increasing $d_n$, the influence of the sensing target conditions will be more significant.
Nevertheless, to obtain a precise {\rcsf}, numerical full-wave simulation and practical experiments are in need, which is of high computational time.
Therefore, to reduce the time consumption required in optimizing $\bm d$ while ensuring the effectiveness of the results, we use $\hat{\gamma}(f,  \bm c,\bm d)$ and an additional interpolation function together to fit the precise {\rcsf} over a sampled set of $\bm d$ denoted by $\hat{\mathcal D}_\mathrm{A}$.
The resulting model-based fitting function is denoted by ${\gamma}(f,  \bm c,\bm d)$ and used in the following optimization of $\bm d$.

%%%%%%%%%%%%%%%%%%%%%%%%%%
\section{System Model}
%%%%%%%%%%%%%%%%%%%%%%%%%%
\label{sec: system model}

\begin{figure}[!t]
\center{\includegraphics[width=0.9\linewidth]{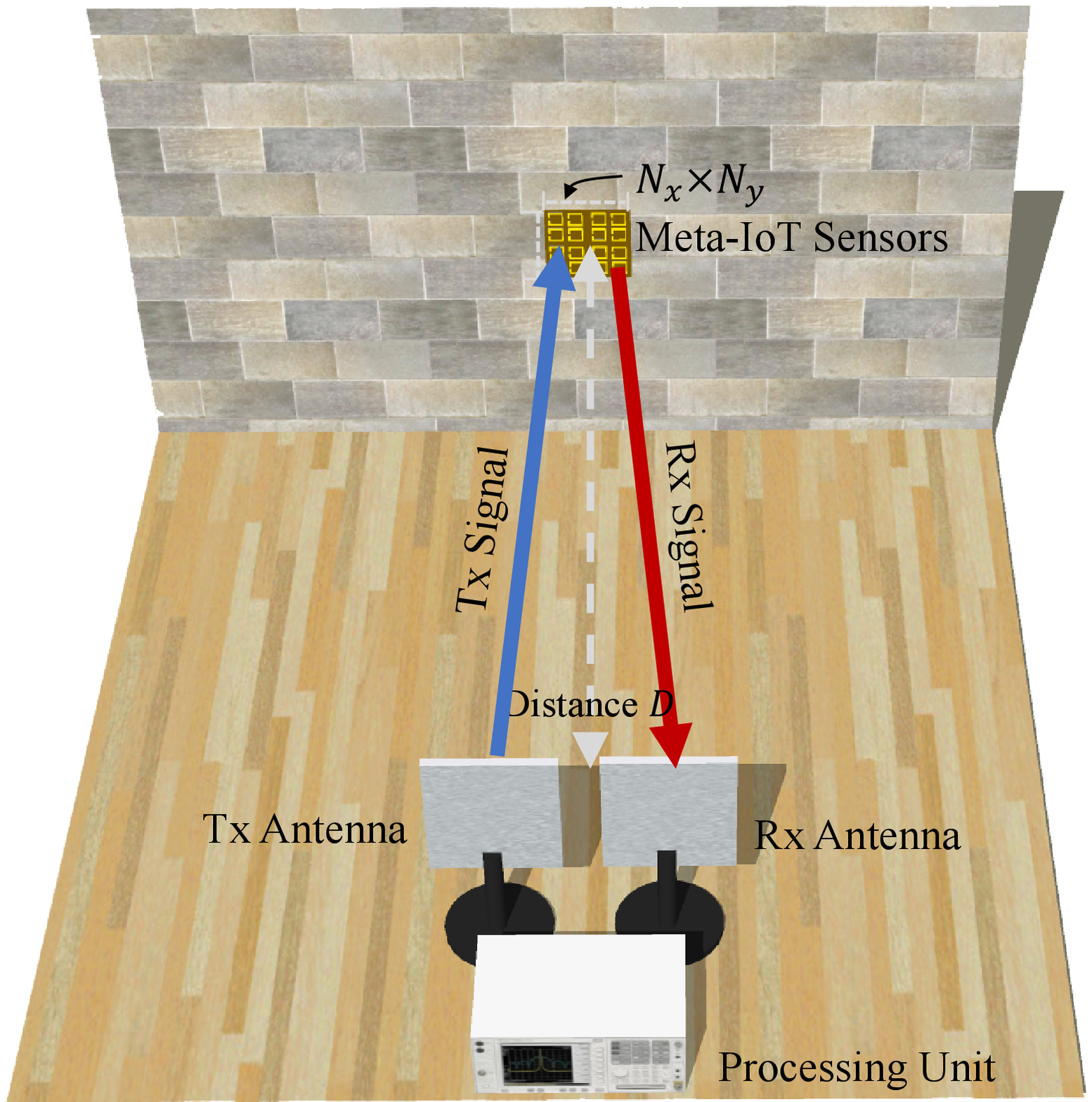}}
%	\vspace{-.3em}
	\setlength{\belowcaptionskip}{-1.em} %调整图片标题与下文距离 
\caption{{\Scenarioname}.}
\vspace{-1.em}
	\label{fig: scenario}
\end{figure}

In this section, we first describe components of the {\scenarioname}, and then establish the transmission model.

%=======================
\subsection{System Description}
%=======================
The {\scenarioname} consists of a wireless transceiver and an array of {\mmr}s, as shown in Fig.~\ref{fig: scenario}.
The {\mmr} array is composed of $N_\mathrm{x} \times N_\mathrm{y}$ {\mmr}s densely paved within a 2D rectangle region, which is fixed on the wall in front of the wireless transceiver.
The wireless transceiver consists of a processing unit and a pair of Tx and Rx antennas, which can be potentially carried by different devices, such as access points, base stations, or unmanned aerial vehicles~\cite{Zhang2019Cellular2}.
The transceiver is capable of transmitting and receiving signals within frequency range $[f_{\mathrm{lb}}, f_{\mathrm{ub}}]$.
The {\rxunit} uses a \emph{{\sensfunc}} denoted by $\bm g$ to map the power of received signals to the sensing target conditions.

%======================
\subsection{Transmission Model}
%======================
% 在这个部分里面描述一下含有超材料单元的信道

Denote the transmit power by $P$, and with the help of~\cite{Vena2014AFully}, we can model the received signal power as
\begin{align}
\label{equ: channel model H}
 &P_{\rx, \dB}(f, \bm d; \bm c) =  10\log_{10}\!\big(\!\underbrace{\PL_{\tx,\rx}(f)}_{\text{Pathloss}}\nonumber\\
& \quad \cdot(\underbrace{\eta_{\mathrm{env}}\!\cdot\! P \!\cdot\! R_{\mathrm{W}}+ \eta_{\mathrm{ms}}\!\cdot\! P\!\cdot\! \gamma(f, \bm c, \bm d)}_{\text{Reflection}}) \!+\! \underbrace{P_\mathrm{b}}_{\text{Bias}} \big)+ \underbrace{e(f)}_{\text{Noise}},
\end{align}
Equation~(\ref{equ: channel model H}) consists of four parts: pathloss, reflection, bias, and noise, which are explained in detail as follows.

%----------------------------
\subsubsection{Pathloss}
%----------------------------
Based on~\cite{goldsmith2005wireless}, the pathloss can be modeled~by 
	\begin{equation}
	\label{equ: tx2rx pathloss factor}
		\PL_{\tx, \rx}(f)=  (\frac{v}{4\pi f})^2\cdot (\frac{1}{2D})^\alpha,
	\end{equation}
where $\alpha$ is the path loss index, and $D$ denotes the distance between the antenna and the {\mmr}, i.e., the \emph{measurement distance}.

%--------------------------------------------
\subsubsection{Reflection}
%--------------------------------------------
The reflection part is composed of two terms.
The first term, i.e., $\eta_{\mathrm{env}}\!\cdot P\!\cdot R_{\mathrm{W}}$, is the power reflected by the wall, where $R_{\mathrm{W}}$ denotes the reflection coefficient of the wall and can be obtained with the help of~\cite{Landron1996AComparison}.
Besides, the second term, i.e., $\eta_{\mathrm{ms}}\cdot P \cdot \gamma(f,  \bm c, \bm d) $ indicates the power reflected by the {\mmr}s.
Here, $\eta_{\mathrm{ms}}$ and $\eta_{\mathrm{env}}$ can be calculated by
\beq
\label{equ: eta related on d}
\eta_{\mathrm{ms}} =\frac{A}{S_0\cdot(D/D_0)^2}, ~\eta_{\mathrm{env}} = 1 - \eta_{\mathrm{ms}},
\eeq
where $S_0$ denotes the coverage area of the antenna's radiation on the plane at a unit distance, and $A = N_\mathrm{T} l_\mathrm{s}^2 N_\mathrm{x} N_\mathrm{y}$ is the area of the {\mmr} array shown in Fig.~\ref{fig: scenario}.

%----------------------------------------
\subsubsection{Bias}
%----------------------------------------
The bias at the wireless receiver accounts for the ambient environmental signals and the power measurement bias of the receiver.
The bias $P_\mathrm{b}$ is modeled as a constant value that is much smaller than the transmitted power $P$.

%----------------------------------------
\subsubsection{Noise}
%----------------------------------------
We model measurement noise $e(f)$ of the wireless transceiver as a random variable following Gaussian distribution $\mathcal N(0, \sigma_{\mathrm{M}}^2)$, where $\sigma^2_{\mathrm{M}}$ denotes the variance of the measurement noise.

%The noise power at a certain frequency $f$ can be modeled as the square of a complex Gaussian noise signal with zero mean and $\sigma_{n}(f)^2/2$ variance, i.e., $p_n(f) = n_r(f)^2 + n_i(f)^2$, where $n_r(f)$ and $n_i(f)$ are the real and imaginary parts of the complex noise signal, respectively, and $n_r(f), n_i(f)\sim \mathcal N(0, \sigma_{n}(f)^2/2)$.
%Therefore, based on~\cite{}, after normalization, $\sqrt{2} n(f)/sigma_{n}(f)$ follows a chi-square distribution with $2$ degrees of freedom, i.e., $\sqrt{2} n(f)/\sigma_{n}(f) \sim \chi_2^2$.
% 取对数
%% 下面这部分好像除了在定理、附录里面之外都没有用上, 所以这里我们就把它删了, 只留在附录里面.
%In summary, by substituting (\ref{equ: tx2rx pathloss factor}) into (\ref{equ: channel model H}), we can express $P_{\rx}(f, d_T, d_H; t, h)$ in dB as follows,
%\begin{equation}
%\label{equ: reformulated p-rx}
%P_{\rx, dB}(f, d_T, d_H; t, h) = \Delta\psi_{dB}(f)  + \gamma_{dB}(f, d_T, d_H; t, h),
%\end{equation}
%where $ \gamma_{dB}(f, d_T, d_H; t, h) = \mu_{dB}(f) + \log_{10}(\eta_{\mathrm{env}}\cdot P \cdot R_{wall}+ \eta_{ms}\cdot P\cdot \tilde{\sigma}(f, d_T, d_H; t, h)) $ indicates the noise-independent.

%%%%%%%%%%%%%%%
\section{Joint {\MmRD} and {\SensFunc} Optimization Problem}
%%%%%%%%%%%%%%%
\label{sec: problem formulation}

In this section, we formulate a joint {\mmrd} and {\sensfunc} optimization problem for {\mmr}s to optimize the sensing performance of the {\scenarioname}, where the influence of both the sensing and transmission is considered jointly to minimize sensing errors.

To evaluate the sensing error, we adopt the root mean squared error~(RMSE) as the loss function.
Besides, the optimization variables are the {\mmrd}, i.e., $\bm d$, and the parameters of the {\sensfunc}, which are coupled together in the formulated optimization problem.
Specifically, the joint {\mmrd} and {\sensfunc} optimization problem is formulated as
\begin{align}
\text{(P1):}\min_{\bm w, \bm d} & ~L_{\mathrm{RMSE}}(\bm w, \bm d) = 
\big(
\sum_{j=1}^{N_\mathrm{C}}
\sum_{m=1}^{N_\mathrm{M}}
\sum_{n=1}^{N_\mathrm{T}}
{\|\tilde{\bm c}_{j,m} - \bm c_{j}\|_2^2\over N_\mathrm{C}N_\mathrm{M}N_\mathrm{T}} \nonumber
\big)^{1/2}, \\
% constraint next
\label{equ: p1 const 1}
s.t.~ & (\bm p_{j,m}, {\bm c}_j) \!\in\! \mathcal D,~\forall j\!\in\![1, N_\mathrm{C}], m\!\in\![1,N_\mathrm{M}], \\
% constraint next
\label{equ: p1 const 2}
& \tilde{\bm c}_{j,m} = \bm g^{\bm w}(\bm p_{j,m}), ~\forall j\!\in\![1, N_\mathrm{C}], m\!\in\![1,N_\mathrm{M}],\\
% constraint next
\label{equ: p1 const 3}
& \bm p_{j,m}\! = \!\left(P_{\rx, \dB}^{(m)}\!(f_1, \bm d; \bm c_j), ..., P_{\rx, \dB}^{(m)}\!(f_{N_\mathrm{F}}, \bm d; \bm c_j)\right)\!, \nonumber \\
&\hspace{8em} j\in[1, N_\mathrm{C}],m\in[1,N_\mathrm{M}], \\
% constraint next
\label{equ: p1 const 4}
& P_{\rx,\dB}^{(m)}\!(f_i, \bm d; \bm c_j) =10\log_{10}\big(\PL_{\tx,\rx}(f_i)\\ 
&\cdot(\eta_{\mathrm{env}}\!\cdot\! P \!\cdot \!R_{\mathrm{W}}\!+\! \eta_{\mathrm{ms}}\!\cdot\! P\!\cdot\!\gamma(f_i, \bm c_j, \bm d ))\!+\!P_\mathrm{b}\big)\!+\!e_{i}^{(m)},\nonumber\\
&\hspace{5em} i\in[1,N_\mathrm{F}],m\in[1,N_\mathrm{M}], j\in[1, N_\mathrm{C}],\nonumber \\
% constraint next
\label{equ: p1 const 5}
& \bm d \in \mathcal D_\mathrm{A},
\end{align}
% 这里要一个一个地解释
where $N_\mathrm{F}$ denotes the number of measured frequencies.

In the objective function of (P1), $N_\mathrm{C}$ is the number of considered sensing condition vectors, $N_\mathrm{M}$ denotes the number of measurements given each sensing target condition vector, $\bm c_j$ indicates the $j$-th considered sensing target condition vector, and $\tilde{\bm c}_{j,m}$ is the estimated sensing target condition vector for the $m$-th measurement at $\bm c_j$.
%where $\mathcal D_A$ denotes the set of available {\mmrd} vectors, and $e_i\sim\mathcal N(0, \sigma_m)$ is the Gaussian measurement noise at frequency $f_i$ with $\sigma_m$ being the variance.
Constraint~(\ref{equ: p1 const 1}) indicates that the $N_\mathrm{M} N_\mathrm{C}$ measurements given $N_\mathrm{C}$ sensing target condition vectors constitute the data set to minimize the RMSE of the system, where $\bm p_{j,m}$ denotes the $m$-th {\emph{\feavec}} at $\bm c_j$.
Constraint~(\ref{equ: p1 const 2}) is due to that the sensing target condition vector is estimated by using $\bm g^{\bm w}$.
Constraints~(\ref{equ: p1 const 3}) and (\ref{equ: p1 const 4}) indicate the received power vector is determined by the wireless propagation channel, which is influenced by the sensing targets and the {\mmrd}.
Specifically, superscript $(m)$ denotes the result of the $m$-th measurement.
Besides, in (\ref{equ: p1 const 5}), $\mathcal D_\mathrm{A}$ denotes the set of available {\mmrd} vectors, which is a continuous set.
%In (P1), the influence of the sensing and transmission is jointly considered by constraint (\ref{equ: p1 const 4}), and the {\sensfunc} can handle this influence through solving~(P1).

Nevertheless, due to the {\ist} of the {\mmr}, the {\mmrd}, i.e., $\bm d$, impacts the sensing and transmission jointly, which makes (P1) a challenging problem.
Moreover, constraint (\ref{equ: p1 const 4}) is highly non-convex due to the reflection coefficient function and is hard to evaluate.
To solve~(P1) efficiently, we decompose it into two sub-problems and propose the algorithms to solve them sequentially in Section~\ref{sec: alg design}.

%%%%%%%%%%%%%%
\section{Algorithm Design}
%%%%%%%%%%%%%%
\label{sec: alg design}
In this section, we propose an efficient algorithm to solve the formulated problem (P1).
In~(P1), the main challenges lie in $\bm d$ affects both {\mmr}s' sensitivity towards the sensing targets and the transmission, and optimization of $\bm d$ and $\bm w$ being coupled.
To handle these challenges, we decompose (P1) into two subproblems, i.e., \emph{{\mmrd} optimization}, and \emph{{\sensfunc} optimization}, and solve them sequentially.

%===================================
\subsubsection{{\MmRD} Optimization Algorithm}
%===================================
The {\mmrd} optimization is based on the following intuition.
Consider $\bm g^{\bm w}$ as a general classification function, which recognizes the sensing target conditions corresponding to a {\feavec}.
Then, to optimize the potential performance of $\bm g^{\bm w}$ requires to minimize the \emph{indiscernibility} of the {\feavec}s for different sensing target condition vectors.

% 不同温度产生的接收信号向量的可分性可以通过信号向量的相似度
One widely adopted indiscernibility measurement is the \emph{Euclidean distance}, which is used to evaluate the extent that two vectors can be discerned from each other~\cite{goodfellow2016deep}.
For example, the average negative Euclidean distance can be used to measure the indiscernibility of the {\feavec}s in the meta-IoT system, which can be calculated by
\begin{equation}
\label{equ: euler distance}
	I_{\mathrm{ED}} = - \frac{1}{N_\mathrm{C}}\sum_{j,j'\in[1,N_\mathrm{C}]} \big\| \hat{\bm p}_j - \hat{ \bm p}_{j'}\big\|_2^2,
\end{equation}
where $\hat{\bm p}_j$ denotes the expectation of $\bm p_{j, m}$, i.e., neglecting the influence of measurement noise.

Nevertheless, using $I_{\mathrm{ED}}$ to measure the indiscernibility is inaccurate and inefficient for the considered meta-IoT sensing system.
% 具有较大差异的气候条件会导致 \feavec 差异也很大, 不容易认错.
Intuitively, two sensing target condition vectors which are largely different from each other will result in significant differences between their corresponding {\feavec}s, which makes them highly distinguishable. 
Thus, the indiscernibility should be majorly due to the {\feavec}s of neighboring sensing target condition vectors which are similar to each other.
Besides, calculating $I_{\mathrm{ED}}$ in~(\ref{equ: euler distance}) is of the computational complexity $\mathcal O(N_\mathrm{C}^2)$, which is time-consuming when $N_\mathrm{C}$ is large.
To handle the above issues, we adopt \emph{the error probability for judging nearest neighbors} as the indiscernibility measurement for the {\feavec}s, which is
\begin{equation}
\label{equ: proposed indiscernibility measurement}
I_{\mathrm{EN}} = \sum_{j=1}^{N_\mathrm{C}}\sum_{{j'}\in\mathcal N_{j}} \Pr^{\err}(\bm c_{j'}| \bm c_j),
\end{equation}
where $\mathcal N_{j}$ denotes the index set of nearest neighbors of the $j$-th sensing target condition vector, 
and $\Pr^{\err}(\bm c_{j'}| \bm c_j)$ denotes the error probability to judge the $N_\mathrm{T}$ sensing target conditions to be $\bm c_{j'}$ when the actual conditions are $\bm c_{j}$.
Specifically, nearest neighbor set $\mathcal N_j$ in (\ref{equ: proposed indiscernibility measurement}) is composed of the sensing target condition vectors which have the smallest positive or negative difference with the $j$-th sensing target condition vector at each sensing target condition, i.e.,
\begin{align}
\label{equ: nearest neighbor set}
&\mathcal N_j = \cup_{n=1}^{N_\mathrm{T}}
\Big\{j'| j'\!=\!\argmin_{j''\in\mathcal X} \sum_{n'\neq n}|c_{j'',n'} - c_{j,n'}|,  \\
&\hspace{.5em} \text{ where } \mathcal X\!=\!\{ \argmin_{\rho\in[1,N_\mathrm{C}],\rho\neq n} |c_{\rho,n} - c_{j,n}|\text{ and } c_{\rho,n}\! \neq \! c_{j,n}\}
\Big\},\nonumber
\end{align}
where $c_{j,n}$ indicates the $n$-th element of vector $\bm c$.
It can be observe in~(\ref{equ: nearest neighbor set}) that $|\mathcal N_j|\leq 2N_\mathrm{T}$.
Therefore, the computational complexity of $I_{\mathrm{EN}}$ is $\mathcal O(N_\mathrm{C})$, which is much smaller than that of $I_{\mathrm{ED}}$.
Besides, the error probability $\Pr^{\err}(\bm c_{j'}|\bm c_j	)$ can be calculated with the help of Proposition~\ref{prop: distinguish ability}.

\begin{proposition}
\label{prop: distinguish ability}
Assume that the maximum likelihood decision criterion is adopted to judge between the sensing target condition vectors.
Then, for a given {\mmrd} $\bm d$, $\Pr^{\err}(\bm c_{j'}|\bm c_j)$ can be calculated by
%前文中需要仔细定义一下, 接收信号序列怎么说, 然后经过log处理的接收信号序列怎么说, 然后排除噪声的接收信号序列怎么说.
\begin{equation}
\label{equ: pr err}
\Pr^{\err}(\bm c_{j'}|\bm c_j) = 0.5\cdot \big(1-\erf(\frac{\sum_{i=1}^{N_\mathrm{F}} {(\tau_{j',i} - \tau_{j,i} )^2}}{2\sqrt{2}}) \big),
\end{equation}
where $\erf(\cdot)$ denotes the \emph{error function}, and $\tau_{j,i}$ ($\tau_{j',i}$) is
\begin{align}
\label{equ: gamma ji}
&\tau_{j,i}\! =\! 10\log_{10}\!\big(\PL_{\tx,\rx}(f_i)\!\cdot\! P\!\cdot\!(\eta_{\mathrm{env}}R_{\mathrm{W}} \!+\! \eta_{\mathrm{ms}}\gamma(f_i, \bm c_j, \bm d )) \nonumber \\
&\hspace{6.em}+\!P_\mathrm{b}\big). 
\end{align}
Moreover, $\Pr^{\err}(\bm c_{j'}|\bm c_j)$ decreases as transmit power $P$ increases or as measurement distance $D$ decreases.
\end{proposition}
\begin{IEEEproof}
See Appendix~\ref{appx: derivation of weighted error prob}.	
\end{IEEEproof}

Therefore, based on~(P1) and (\ref{equ: proposed indiscernibility measurement}), the {\mmrd} optimization problem can be formulated as follows, where we minimize $I_{\mathrm{EN}}$ by optimizing $\bm d$.
\manuallabel{opt: sensor design optimization}{sP1}
\begin{align}
\text{(sP1):} \min_{\bm d} ~%
&I_{\mathrm{EN}} = \sum_{j, j'\in[1, N_\mathrm{C}]}\Pr^{\err}(\bm c_{j'}|\bm c_j),  \nonumber\\
s.t. ~ &\text{(\ref{equ: p1 const 5}), (\ref{equ: nearest neighbor set})-(\ref{equ: gamma ji}).} \nonumber 
\end{align}
Due to that the objective function in (\ref{opt: sensor design optimization}) contains a non-convex function, i.e., $\erf(\cdot)$, (\ref{opt: sensor design optimization}) is a non-convex optimization problem that is hard to solve.
%Besides, to evaluate the objective function in (\ref{opt: sensor design optimization}), it is necessary to calculate $\tilde{\sigma}_{ms}(f_i, \bm d; \bm c_j)$ by numerical simulation which can be of high time-complexity.
To solve (\ref{opt: sensor design optimization}) efficiently, we adopt the surrogate optimization algorithm~\cite{Wang2014AGeneral}, which can handle finitely bounded non-convex optimization problems and has a high probability of finding a global optimum.

%=====================================
\subsubsection{{\SensFunc} Optimization Algorithm}
%=====================================
\label{ssub: alg for sensing func opt}

In the {\sensfunc} optimization, we adopt the optimized {\mmrd}, i.e., $\bm d^*$, and minimize the RMSE of the sensing by optimizing $\bm w$, i.e.,
\manuallabel{opt: sensing func optimization}{sP2}
\begin{align}
\text{(sP2):}\min_{\bm w} & ~L_{\mathrm{RMSE}}(\bm w, \bm d^*) \!=\! 
\big(\!\sum_{(\bm p_{j,m}, \bm c_j)\in \mathcal D}\! 
\frac{\|\tilde{\bm c}_{j,m}\!-\!\bm c_{j}\|_2^2}{|\mathcal D|}\big)^{1/2}_, \nonumber\\
s.t. &~\text{(\ref{equ: p1 const 1})-(\ref{equ: p1 const 3})}. \nonumber
\end{align}

To solve (\ref{opt: sensing func optimization}), we model that $\bm g^{\bm w}$ as a fully connected neural network, which is an efficient model for general types of classification and regression functions.
The fully connected neural network consists of an input layer, a hidden layer, and an output layer, which are connected successively.
The input layer takes the $N_\mathrm{F}$-dimensional real-valued {\feavec}, passes it to the hidden layer which contains $N_\mathrm{W}$ neural nodes, each of which calculates a biased weighted sum of its input and processes the sum with a \emph{softmax} function~\cite{goodfellow2016deep}.
Then, the nodes in the hidden layer pass the results to the output layer, which consists of $N_\mathrm{T}$ neural nodes, which output the estimated sensing target conditions.
In this case, the parameter vector of the {\sensfunc}, i.e., $\bm w$, stands for the weights of the connections and the biases of the nodes.

\begin{algorithm}[!t]
\small
  \caption{Joint {\MmRD} and {\SensFunc} Optimization Problem}
\hspace*{0.02in} {\bf Input:} %算法的输入，
%\hspace*{0.02in}用来控制位置，同时利用 \\ 进行换行
% set of climate condition
Set of $N_\mathrm{C}$ sensing target conditions; 
Set of $N_\mathrm{F}$ frequency points;
Number of training data $|\mathcal D|$;
Set of available {\mmrd} vectors $\mathcal D_\mathrm{A}$;
Channel coefficients, $\alpha$, $\sigma_{\mathrm{M}}$, $\eta_{\mathrm{env}}$, $\eta_{\mathrm{ms}}$, $P$, and $D$;
Learning rate $\beta$. 
\\
\hspace*{0.02in} {\bf Output:} %算法的结果输出
$\bm d^*$,
$\bm g^{\bm w*}$.
\label{alg: summary algorithm}
\begin{algorithmic} [1]
\State Solve (\ref{opt: sensor design optimization}) and obtain the optimal {\mmrd} vector $\bm d^*$ by using the surrogate optimization algorithm.
\State Based on $\bm d^*$, use Monte Carlo method to generate random set of training data $\mathcal D$.
\State Using the training data set $\mathcal D$ to train the neural network by solving (\ref{opt: sensing func optimization}), and obtain the optimized {\sensfunc} $\bm g^{\bm w*}$.
\State \Return $\bm d^*$ and $\bm g^{\bm w*}$.
\end{algorithmic}
\end{algorithm}

Moreover, to obtain training data set $\mathcal D$ in~(\ref{equ: p1 const 1}), we use the Monte Carlo method.
% 这里说了simulation 里面 training data set 怎么获取, 但是没有说 experiment 里面 training data set 怎么获取.
In the simulation, we generate a set of random {\feavec}s satisfying the constraints in (\ref{opt: sensing func optimization}) which constitutes $\mathcal D$. i.e.,
\begin{align}
\label{equ: training data gen by mc}
\mathcal D =& \big\{ (\bm p_{j, m}, \bm c_j) \big | \bm p_{j, m} \!=\! \bm e \!+\! \bm \tau^*(\bm c_j), \bm e\!=\!(e_1,\dots,e_{N_\mathrm{F}}), \\
&e_i\sim \mathcal N(0, \sigma^2_\mathrm{M}), i\!\in\![1,N_\mathrm{F}], j\!\in\![1, N_\mathrm{C}],  m \!\in\! [1, N_\mathrm{M}] \big\},\nonumber
\end{align}
where 
$\bm \tau^*(\bm c_j)$ is a $N_F$-dimensional vector with its $i$-th element being $\tau^*_{i}(\bm c_j) =10\log_{10}\big(\PL_{\tx,\rx}(f_i)\cdot(\eta_{\mathrm{env}}\cdot P \cdot R_\mathrm{W}+ \eta_{\mathrm{ms}}\cdot P\cdot \gamma(f_i, \bm c_j, \bm d^* )) + P_\mathrm{b}\big)$,
and $\bm e$ is a $N_\mathrm{F}$-dimensional vector of independently distributed Gaussian random variables.

Then, we optimize $\bm g^{\bm w}$ by training it on $\mathcal D$ using the supervised learning technique~\cite{goodfellow2016deep}.
The training of ${\bm w}$ is performed by iteratively updating $\bm w$ along the negative gradient of the RMSE loss in~(\ref{opt: sensing func optimization}), i.e.,
\beq
\label{equ: w update}
\bm w = \bm w - \beta\nabla_{\bm w} L_{\mathrm{RMSE}}(\bm w, \bm d^*),
\eeq
where the gradient $\nabla_{\bm w} L_{\mathrm{RMSE}}(\bm w, \bm d^*)$ is calculated by using the back-propagation algorithm~\cite{goodfellow2016deep}, and $\beta\in [0,1]$ denotes the learning rate.
Sum up the algorithms to solve sub-problems (\ref{opt: sensor design optimization}) and (\ref{opt: sensing func optimization}), and we can summarize the complete algorithm to solve (P1) as Algorithm~\ref{alg: summary algorithm}.

%%%%%%%%%%%%%%%%
\section{Simulation Results}
%%%%%%%%%%%%%%%%
\label{sec: simulation result}

%In this section, to validate the effectiveness of the proposed framework and design technique, we design a {\mmr} sensing system to sense the temperature and humidity levels.
%We first provide simulation results of the proposed algorithm, where the optimal {\mmrd} is obtained.
In this section, we provide simulation results of the  proposed algorithm, which validates the effectiveness of the algorithm and the meta-IoT system.
Besides, we give insight into how the transmit power and the measurement distance influence the sensing accuracy.

\begin{table}
\centering
	\caption{Simulation Parameters}	
	\label{table: simulation parameters}
\includegraphics[width=0.95\linewidth]{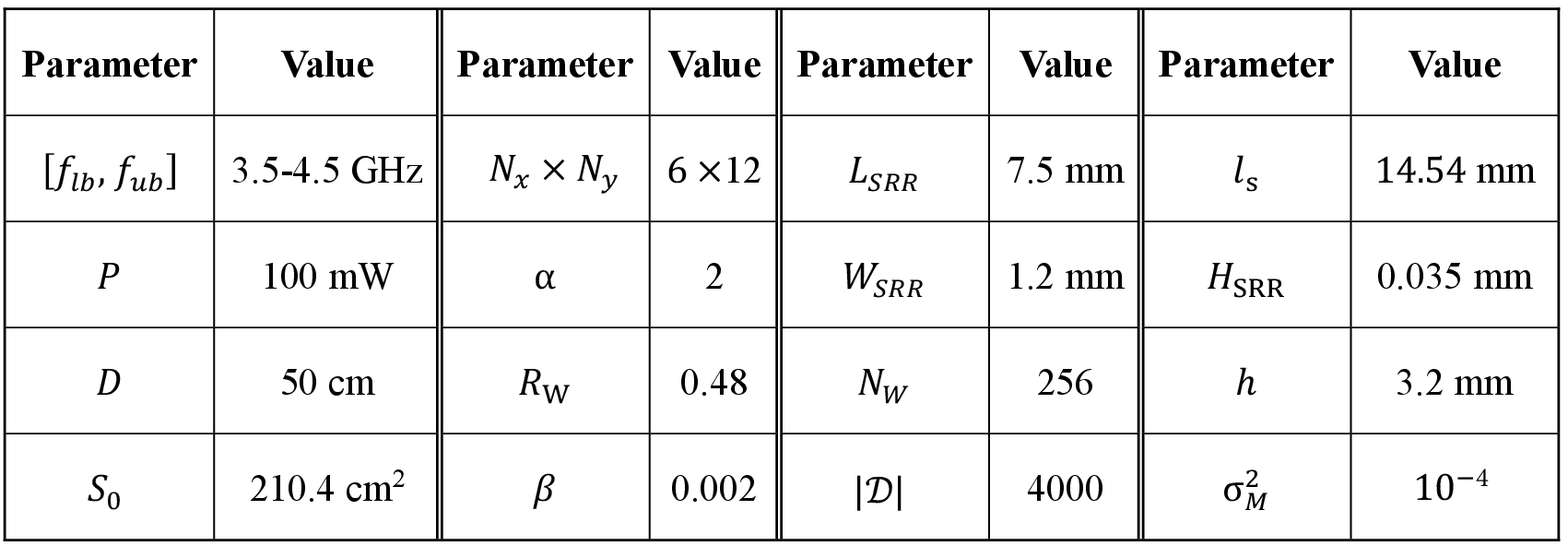}
	\vspace{-1em}
\end{table}

In the simulation, each {\mmr} consists of two {\mmu}s.
The first {\mmu} has {\tsm} within its gap and is aimed for sensing temperature, which we refer to as the \emph{T-unit}.
Similarly, the second {\mmu} contains {\hsm} for sensing humidity and is referred to as \emph{H-unit}.
More specifically, the {\mmu}s are made of copper rings and FR-4 supportive substrate, and the detailed setting parameters are shown in Table~\ref{table: simulation parameters}.
The adopted {\tsm} in T-unit is the powder used in the NTC thermistor SDNT2012X102-3450-TF~\cite{Tmaterial}.
Besides, the adopted humidity-sensitive material in H-unit is the polymer used in the hygristor TELAiRE HS30P~\cite{Hmaterial}.
Moreover, the selected sensing target condition set is $\mathcal C = \{(c_1, c_2)| c_1\in[5,45], c_2\in[20,60], c_1,c_2\equiv 0 \mod 5\}$, and thus $N_C = 81$.
By solving (P1) given the above simulation settings, we obtain the optimal {\mmrd} $\bm d^* = (2.05, 1.22)$ mm.

%@@@@@@@@@@@@@@@@@@@@@@@@@@
\begin{figure}[t]
\centering
  \includegraphics[width=0.75\linewidth]{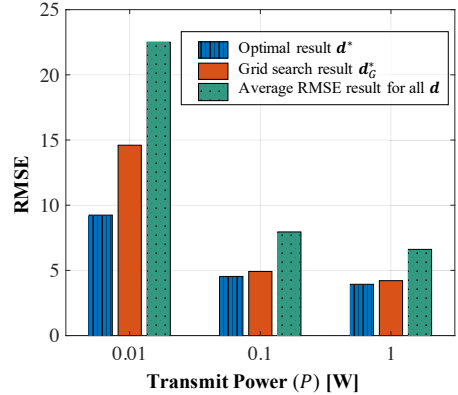}
	\vspace{-.1em}
  \caption{%
 Resulting RMSEs of the meta-IoT system versus transmit power $P$ given different {\mmrd}s.
  }%
     \vspace{-1.2em}
  \label{fig: P influence}
\end{figure}
\begin{figure}[t]
\centering
  \includegraphics[width=0.75\linewidth]{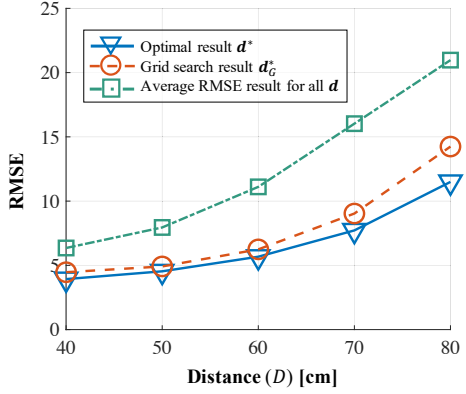}
	\vspace{-.1em}
  \caption{%
 RMSEs of the meta-IoT system versus distance between the wireless transceiver and the {\mmr} given different {\mmrd}s.
  }%
     \vspace{-1.2em}
  \label{fig: D influence}
\end{figure}
%@@@@@@@@@@@@@@@@@@@@@@@@@@

Figs.~\ref{fig: P influence} and~\ref{fig: D influence} show the resulting RMSE obtained by solving (sP2) under different cases of {\mmrd}. 
% 条形的 蓝色 bar 代表着采用经过优化后最优的结果
% 红色的 bar 代表着从粗粒度的枚举中最优结果, 也就是 在 上面那个图里面的 d = (4,1) 得到的结果
% 点状的 绿色 bar 
% 查一下填充是用哪个词来形容. 代表的是网格的d的平均结果.
In both Figs.~\ref{fig: P influence} and~\ref{fig: D influence}, the first case indicates the resulting RMSE of the meta-IoT system with the optimal {\mmrd} found within $\mathcal D_\mathrm{A} = \{\bm d\in\mathbb R^2 | d_\mathrm{T}, d_\mathrm{H}\in [1,5] \text{ mm}, d_\mathrm{T}\neq d_\mathrm{H}\}$.
The second case indicates the resulting RMSE of the meta-IoT system given the optimal integer {\mmrd} found within $\hat{\mathcal D}_\mathrm{A} = \{\bm d\in\mathbb Z^2 | d_\mathrm{T}, d_\mathrm{H}\in [1,5] \text{ mm}, d_\mathrm{T}\neq d_\mathrm{H}\}$, which is denoted by $\bm d_G^*$.
The third case indicates the average resulting RMSE of the meta-IoT system given {\mmrd}s in $\hat{\mathcal D}_\mathrm{A}$.
It can be observed that the optimal {\mmrd} obtained by using Algorithm~1 outperforms the other {\mmrd}s in terms of the resulting RMSE.

Besides, it can be observed that the resulting RMSE values of different {\mmrd}s decrease as the transmit power increases, and increase as the measurement distance increases, which are in accordance with the analysis in Proposition~\ref{prop: distinguish ability}.
Moreover, the optimal {\mmrd} leads to the lowest RMSE under different transmit power and distance.

%%%%%%%%%%%%%
\section{Conclusion}
%%%%%%%%%%%%%
\label{sec: conclu}

In this paper, we have designed a {\scenarioname}, which is composed of {\mmr}s and a wireless transceiver.
In the meta-IoT system, the {\mmr}s can simultaneously sense physical environmental conditions and transmit back the sensing results in a fully passive manner.
We have analyzed the relationship between the {\mmrd} and its reflection coefficients for wireless signals as well as the transmission model. 
A joint {\mmrd} and {\sensfunc} optimization problem has been formulated and solved, in which the {\mmrd}’s influence on the sensing and transmission has been jointly considered.
Specifically, we have applied the proposed {\scenarioname} to sense the temperature and humidity levels.
Simulation results have verified the effectiveness of the proposed algorithm in minimizing the RMSE of {\mmr}’s sensing for temperature and humidity levels, and have revealed the trend that the sensing accuracy decreases as the transmit power decreases or the measurement distance increases.

%\vspace{-.6em}
\begin{appendices}
\section{Proof of Proposition~\ref{prop: distinguish ability}}
\label{appx: derivation of weighted error prob}
	\vspace{-.5em}
% $\hat{p}_{j, i} $ 是排除了随机因素的 anchor 向量, 也可以称为 特征向量

Given $\bm c_{j}$~($\bm c_{j'}$), we denote the corresponding received power vector by ${\bm p}_{j}$~(${\bm p}_{j'}$).
In the following, we analyze the probability to decide the sensing target conditions to be $\bm c_{j'}$ by judging from ${\bm p}_{j}$.
Based on~(\ref{equ: channel model H}) and~(\ref{equ: gamma ji}), the $i$-th elements of $\bm p_{j}$ and $\bm p_{j'}$ are
\begin{equation}
	p_{j, i} = e_{i} + \tau_{j, i},~	p_{j', i} = e_{i} + \tau_{j', i},
\end{equation}
where $ e_i$ is a random variable with random normal distribution $\mathcal N(0, \sigma_{\mathrm{M}}^2)$ with $\sigma_{\mathrm{M}}^2$ being the variance of the measurement noise at frequency $f_i$, 
and $\tau_{j, i}$ is defined in (\ref{equ: gamma ji}).
Then, based on the probability density function of the normal distribution, the probability of getting $\bm p_{j}$ given $\bm c_{j}$ and $\bm c_{j'}$ are
\begin{align}
&\Pr( \bm p_{j}|\bm c_{j}) \!=\! \frac{\prod_{i=1}^{N_\mathrm{F}}e^{-\frac{e_i^2}{2\sigma_{\mathrm{M}}^2}}}{\sqrt{2\pi}\sigma_{\mathrm{M}}}\!,
\Pr( \bm p_{j}|\bm c_{j'}) \!=\! \frac{\prod_{i=1}^{N_\mathrm{F}}\!e^{-\!\frac{(e_i \!+\! \tau_{j,i} \!-\! \tau_{j',i})^2}{2\sigma_{\mathrm{M}}^2}}}{\sqrt{2\pi}\sigma_{\mathrm{M}}} . \nonumber
\end{align}

Using the maximum likelihood criterion, the probability to decide on $\bm c_{j'}$ from $\bm p_j$ can be calculated by $\Pr( \bm p_{j}|\bm c_{j'})  > \Pr( \bm p_{j}|\bm c_{j})$, which is equivalent to
\begin{align}
\label{equ: judging criterion}
%&\Pr\{\sum_{i=1}^{N_F} \frac{\Delta \psi_{dB, i}^2}{\sigma_{dB}^2} > \sum_{i=1}^{N_F} \frac{(\Delta \psi_{dB,i} - (\gamma_{j_2,i} - \gamma_{j_1,i}))^2}{\sigma_{dB}^2} \}\\
%\Leftrightarrow 
%&
\Pr\big\{
\sum_{i=1}^{N_\mathrm{F}} 2{e_i \cdot (\tau_{j', i} - \tau_{j,i})} > \sum_{i=1}^{N_\mathrm{F}} 
{ (\tau_{j',i} - \tau_{j,i})^2} 
\big\}.
\end{align}

Denote $\chi_{j', j} = \sum_{i=1}^{N_\mathrm{F}} {e_i \cdot (\tau_{j', i} - \tau_{j,i})}$, and $\chi_{j', j}$ is a random variable following the normal distribution, i.e., $\chi_{j', j}\sim\mathcal N(0, \sigma^2_{j', j})$, where
\begin{equation*}
\sigma_{j', j}^2=\sigma_\mathrm{M}^2\sum_{i=1}^{N_\mathrm{F}} {(\tau_{j',i} - \tau_{j,i} )^2}.
\end{equation*}
% 目前来说感觉这个推导有一点问题, 不是很符合直觉, 因为信道的shadowing越大, 反而错误概率越低, 这明显是不科学的.

Then, the probability in (\ref{equ: judging criterion}) can be calculated by 
\begin{equation}
\label{equ: perr}
\Pr^{\err}(\bm c_{j'}|\bm c_{j}) = 	0.5 - 0.5\cdot\erf(\frac{\sigma_{j', j}^2}{2\sqrt{2}\sigma^2_\mathrm{M}}).
\end{equation}

By taking partial derivative, it can be proved that $\partial \sigma^2_{j', j} /\partial P > 0$, indicating that the error probability decreases as the transmit power increases.
Similarly, by taking the partial derivative with respect to $D$, it can be proven that $\partial \sigma^2_{j', j} /\partial D < 0$, indicating that the error probability decreases as the measurement distance decreases.$\hfill\blacksquare$\par

 \end{appendices}
 \iffalse
 \section*{Acknowledgement}
This work was supported by the National Natural Science Foundation of China under Grant 61931019, Grant 61625101, Grant 61829101, and Grant 61941101.
%\vspace{-1em}
\section*{Acknowledgement}\vspace{-0.2em}
This work was supported in part by the National Nature Science Foundation of China under grant number 61625101.
\fi
\bibliographystyle{ieeetr}%
\bibliography{main}
\end{document}